\documentclass[12pt]{article}
\usepackage{graphics}
\usepackage{graphicx}
\DeclareGraphicsExtensions{.pdf}
\usepackage{float} 
\usepackage{amsmath}
\usepackage{slashed}
\usepackage{amsfonts}   
\usepackage{amssymb}

\usepackage{color}

\begin{document}
\date{}
\title{{\bf{\Large Covariant action for M5 brane in nonrelativistic M-theory}}}
\author{
 {\bf {\normalsize Dibakar Roychowdhury}$
$\thanks{E-mail:  dibakarphys@gmail.com, dibakar.roychowdhury@ph.iitr.ac.in}}\\
 {\normalsize  Department of Physics, Indian Institute of Technology Roorkee,}\\
  {\normalsize Roorkee 247667, Uttarakhand, India}
\\[0.3cm]
}

\maketitle
\begin{abstract}
We construct the nonrelativistic covariant world-volume action for a single M5 brane of $ D=11 $ supergravity in M-theory. The corresponding non-Lorentzian (NL) background possesses a codimension three foliation and is identified as the Membrane Newton-Cartan manifold in the presence of background fluxes that are suitably expanded in $ 1/c^2 $ expansion. We also expand the associated world-volume fields in $ 1/c^2 $ expansion. The above procedure eventually results into a well defined world-volume action that is coupled to Membrane Newton-Cartan background.
\end{abstract}
\section{Overview and Motivation}
Nonrelativistic (NR) limits of quantum gravity and holography have witnessed a steady progress over the past few decades and specially during the last couple of years. These include the understanding of the string sigma models as well well extended objects like $ p $ brane world-volume theories that are formulated over Newton-Cartan (NC) backgrounds. 

However, in spite of these breakthroughs, our current understanding of NR limits of M theory and the extended objects that live in $ D=11 $ supergravity backgrounds is quite limited. The purpose of the present paper is to fill up some of these gaps in the literature. 

In the literature, there exists two parallel approaches to NR strings \cite{Gomis:2000bd}-\cite{Gomis:2005pg} that propagate over non-Lorentzian (NL) backgrounds. The first of these approaches is based on taking a String Newton-Cartan (SNC) limit \cite{Bergshoeff:2015uaa}-\cite{Bergshoeff:2019pij} of relativistic sigma models. The other approach is based on the null reduction procedure \cite{Harmark:2017rpg}-\cite{Harmark:2018cdl} of Lorentzian manifolds\footnote{See \cite{Gomis:2020fui}-\cite{Hartong:2021ekg} for related works.}. Recently, SNC framework has been extended to include background NS-NS fluxes where the presence of the B field is taken care of by additional symmetry generators that eventually leads to F-String Galilei (FSG) algebra \cite{Bidussi:2021ujm}-\cite{Yan:2021lbe}. 

The above idea of SNC limit has been further generalized to study the NR limit of extended objects like $ p $ branes \cite{Roychowdhury:2019qmp}-\cite{Bergshoeff:2022pzk} that are coupled to $ (p+1) $ forms. These constructions could be further uplifted to study membrane solutions in the NR limits of M theory where the starting point should be the codimension three foliation of $ D=11 $ supergravity target space \cite{Ebert:2021mfu}-\cite{Blair:2021waq}. This is precisely the theme of the present paper.

In the present paper, we consider a codimension three foliation of $ D=11 $ M theory background while the subleading corrections to the metric appears at an order $ \mathcal{O}(c^{-1}) $. On top of this, we also expand the world-volume fields ($ X^\mu $) using a $ 1/c^2 $ expansion \cite{Hartong:2021ekg}. When put together, these result into a nice $ 1/c $ expansion in the NR limit of M5 branes \cite{Pasti:1997gx}-\cite{Lambert:2020scy} that are coupled to a Membrane Newton-Cartan (MNC) background.

The organization for the rest of the paper is as follows. In Section 2, we summarise the M5 brane world-volume theory. The NR expansion is carried out in Section 3. Finally, we draw our conclusion and outline some possible future directions in Section 4.
\section{M5 brane world-volume theory}
\label{sec2}
We set the stage by introducing the relativistic M5 brane world-volume theory. The bosonic sector of the M5 brane world-volume theory comprises of a rank two self dual world-volume two form ($ B^{(2)} $) coupled with five world-volume scalars. The vev of these scalars define the location of the M5 brane along transverse axes. 

The abelian M5 brane world-volume action in $ D=11 $ is given by \cite{Bandos:2000az}
 \begin{eqnarray}
 \label{2.1}
 S_{M5}=T_{M5}\int d^6 \sigma \left[\sqrt{-\det (\hat{g}_{ab} + i H^{\ast}_{ab})} -\frac{\sqrt{-\det \hat{g}_{ab}}}{4 \partial_a \phi \partial^a \phi}\partial_l \phi H^{\ast lmn}H_{mnp}\partial^p \phi \right]+S_{WZ},
 \end{eqnarray}
where $ S_{WZ}  $ is the WZ term associated with M5 brane in $ D=11 $ supergravity \cite{Bandos:1997ui}-\cite{Bandos:2000az}
 \begin{eqnarray}
 \label{e2}
 S_{WZ} = T_{M5}\int \hat{C}^{(6)}+\frac{1}{2} F^{(3)}\wedge \hat{C}^{(3)}.
 \end{eqnarray}

Here, $ F^{(3)} = dB^{(2)}$ is the world-volume three form flux. On the other hand, $  \hat{C}^{(3)} $ and $  \hat{C}^{(6)} $ are respectively the background three form and six form fluxes.
 
Here, $ a,b =0, \cdots ,5 $ denote the world-volume directions of the M5 brane. On the other hand, $ X^{\mu}(\mu =t, u, v, m, \cdots ) $ are the spacetime directions associated with the eleven dimensional target space which define the induced metric on the world-volume as
\begin{eqnarray}
\label{2.3}
\hat{g}_{ab} = g_{\mu \nu}\partial_a X^{\mu}\partial_b X^{\nu}.
\end{eqnarray}

The M5 brane world-volume action \eqref{2.1} comes with an anti-symmetric three form flux 
\begin{eqnarray}
H^{\ast}_{ ab}=\frac{1}{ \sqrt{-(\partial \phi )^2}}H^{\ast}_{abc}\partial^c \phi ~;~H^{\ast abc}=\frac{1}{3! \sqrt{- \det \hat{g}_{ab}}}\varepsilon^{abclmn}H_{lmn},
\end{eqnarray}
together with an auxiliary scalar ($ \phi $) that defines
\begin{eqnarray}
(\partial \phi )^2= \partial_a \phi \hat{g}^{ab} \partial_b \phi.
\end{eqnarray}

The determinant in (\ref{2.1}) can be expanded as \cite{Pasti:1997gx}
\begin{eqnarray}
\label{2.6}
\sqrt{-\det (\hat{g}_{ab} + i H^{\ast}_{ab})} = \sqrt{-\det \hat{g}_{ab} }\left(1+\frac{1}{2}\text{tr}H^2 +\frac{1}{8} (\text{tr}H^2)^2 -\frac{1}{4}\text{tr}H^4 \right)^{1/2},
\end{eqnarray}
where we define the above entities as
\begin{eqnarray}
\frac{1}{2}\text{tr}H^2 =-\frac{1}{2}H^{\ast}_{ ab}H^{\ast ab}~;~\frac{1}{4}\text{tr}H^4=-\frac{1}{4}H^\ast_{am}H^{\ast m}~_b H^{\ast a n}(H^\ast)_n~^b.
\end{eqnarray}
\section{Expansion of the world-volume action}
We begin by introducing the Membrane Newton-Cartan (MNC) expansion of $ D=11$ supergravity backgrounds in M theory. We consider a codimension three foliation of the $ D=11$ background in which there are three \emph{longitudinal} axes labeled by (Lorentz) flat indices $ A,B ~(0 , 1, 2)$ and eight \emph{transverse} directions labeled by (Euclidean flat) indices $ I,J ~(3, \cdots, 10)$. The Lorentz indices ($ A, B $) are accompanied by a ($ 2+1 $) dimensional Minkowski metric $ \eta_{AB} = \text{diag}(-1, 1, 1)$. On the other hand, the transverse indices ($ I,J $) are accompanied by an eight dimensional Riemannian metric of the form $ \delta_{IJ} $. 

We begin by considering the $ 1/c $ expansion of the eleven dimensional metric
\begin{eqnarray}
\label{3.1}
g_{\mu \nu}= c^2 \tau_{\mu \nu}(X)+c^{-1}h_{\mu \nu}(X)+ \cdots ,
\end{eqnarray} 
which is accompanied by an inverse metric
\begin{align}
g^{\mu \nu}=c h^{\mu \nu}(X)+c^{-2}\tau^{\mu \nu}(X) + \cdots.
\end{align}

Here we define the above entities as
\begin{eqnarray}
\label{10}
\tau_{\mu \nu}= \tau_{\mu}~^A \tau_{\nu}~^B \eta_{AB}~;~h_{\mu \nu}= e_{\mu}~^I e_{\nu}~^J \delta_{IJ}.
\end{eqnarray}

Following the orthonormal property $ g_{\mu \nu}g^{\nu \rho}=\delta_\mu^\rho $, one could further show \cite{Blair:2021waq}
\begin{align}
\tau_{\mu \nu}h^{\nu \rho}=0~;~ \tau_{\mu \nu}\tau^{\nu \rho}+h_{\mu \nu}h^{\nu \rho}=\delta_\mu^\rho~;~h_{\mu \nu}\tau^{\nu \rho}=0.
\end{align}

Here, $ \lbrace \tau_{\mu}~^A ,  h_{\mu \nu}\rbrace $ are the fields propagating over NL background that are collectively denoted as the MNC data in this paper. Here, $ \tau_{\mu}~^A  ~(A=0, 1 , 2)$ are the \emph{longitudinal} vielbeins. On the other hand, $ e_{\mu}~^I ~(I=3, \cdots , 10) $ correspond to \emph{transverse} vielbeins.

The above expansion \eqref{3.1} is accompanied by an world-volume expansion \cite{Harmark:2019upf}, \cite{Hartong:2021ekg}
\begin{eqnarray}
\label{3.3}
X^{\mu}= x^{\mu}+c^{-2}y^{\mu}+\mathcal{O}(c^{-4}).
\end{eqnarray}

In what follows, we choose to work with a particular embedding for the M5 brane world-volume axes in which three of the world-volume directions are aligned along longitudinal axes and the rest are along the transverse axes of the MNC manifold. Namely, we denote three longitudinal target space directions as $x^t$, $x^u$ and $x^v$ such that
\begin{align}
x^t=x^t(\sigma^0)~;~x^u=x^u(\sigma^1)~;~x^v=x^v(\sigma^2),
\end{align}
where $\sigma^{\bar{a}}(\bar{a}=0,1,2)$ are the longitudinal axes of the M5 brane.

On the other hand, for the remaining three transverse directions one sets
\begin{align}
x^m=x^m(\sigma^i),
\end{align}
where $\sigma^i (i=3,4,5)$ are the transverse world-volume directions of the M5 brane.

Combining all these facts together, the world-volume metric components could be systematically expanded as
\begin{align}
\label{e15}
&\hat{g}_{00}=c^2 \tau_{00}(x)+y^{\mu}\partial_{\mu}\tau_{tt}(x)(\partial_0 x^{t})^2 + 2 \tau_{tt}(x)\partial_0 x^t \partial_0 y^t + \mathcal{O}(c^{-2})\\
&\hat{g}_{11}=c^2 \tau_{11}(x)+y^{\mu}\partial_{\mu}\tau_{uu}(x)(\partial_1 x^{u})^2 + 2 \tau_{uu}(x)\partial_1 x^u \partial_1 y^u+ \mathcal{O}(c^{-2})\\
&\hat{g}_{22}=c^2 \tau_{22}(x)+y^{\mu}\partial_{\mu}\tau_{vv}(x)(\partial_2 x^{v})^2 + 2 \tau_{vv}(x)\partial_2 x^v \partial_2 y^v + \mathcal{O}(c^{-2})\\
&\hat{g}_{ij}=c^{-1} h_{mn}(x)\partial_i x^m \partial_j x^n + \mathcal{O}(c^{-3}),
\label{e18}
\end{align}
where we define $\tau_{00}=\tau_{tt}(x)(\partial_0 x^t)^2$ and so on.

Using the expansions \eqref{e15}-\eqref{e18}, it is now straightforward to calculate
\begin{align}
\label{3.12}
&\sqrt{- \det \hat{g}_{ab}}=c^{3/2}\sqrt{-\det \tau_{\bar{a}\bar{b}}^{(3)}(x)}\sqrt{\det h^{(3)}_{ij}(x)}\nonumber\\
&-\frac{1}{2 \sqrt{c}}\frac{\Sigma (x,y)}{\sqrt{-\det \tau_{\bar{a}\bar{b}}^{(3)}(x)}}\sqrt{\det h^{(3)}_{ij}(x)}~;~ \bar{a},\bar{b}=0,1,2
\end{align}
where we denote the above functions as
\begin{align}
&\Sigma (x,y) = \tau_{00}(x)\tau_{11}(x)T_{22}(x,y)+\tau_{00}(x)\tau_{22}(x)T_{11}(x,y)+\tau_{22}(x)\tau_{11}(x)T_{00}(x,y)\\
&T_{00}(x,y)=y^{\mu}\partial_{\mu}\tau_{tt}(x)(\partial_0 x^{t})^2 + 2 \tau_{tt}(x)\partial_0 x^t \partial_0 y^t \\
&T_{11}(x,y)=y^{\mu}\partial_{\mu}\tau_{uu}(x)(\partial_1 x^{u})^2 + 2 \tau_{uu}(x)\partial_1 x^u \partial_1 y^u\\
&T_{22}(x,y)=y^{\mu}\partial_{\mu}\tau_{vv}(x)(\partial_2 x^{v})^2 + 2 \tau_{vv}(x)\partial_2 x^v \partial_2 y^v.
\end{align}
\subsection{Expansion of $ \text{tr}H^2 $}
Next, we consider the NR expansion of the world-volume three form flux
\begin{align}
H^{(3)}=\frac{1}{3!}H_{abc}d \sigma^a \wedge d \sigma^b \wedge d \sigma^c ,
\end{align}
where we propose an expansion for the world-volume three form field strength as\footnote{The $ c^3 $ scaling for longitudinal components is understandable. This is because, each longitudinal direction is accompanied by a one form $ \tau_{\bar{a}}~^A $ and an associated factor of $ c $. For a rank three longitudinal tensor, there are three such fields (corresponding to three different indices) and hence a factor of $ c^3 $. Finally, for transverse metric component $ \hat{g}_{ij} $, which is a rank two tensor, the leading term goes as $ c^{-1} $ (see \eqref{e18}). Following the above line of argument, we therefore propose that a natural scaling for a rank three transverse tensor field should go like $ c^{-3/2} $ and for a rank one tensor it should scale as $ c^{-1/2} $.}
\begin{align}
\label{2.21}
&H^{\parallel}_{\bar{a} \bar{b}\bar{c}} =c^3 \mathcal{T}_{\bar{a} \bar{b}\bar{c}}(x)+c y^\mu \partial_\mu \mathcal{T}_{\bar{a} \bar{b}\bar{c}}(x)+\mathcal{O}(c^{-1})\\
&H^{\perp}_{ijk}=c^{-3/2}\pi_{ijk}(x)+\mathcal{O}(c^{-7/2}),
\end{align}
where $\bar{a}, \bar{b}, \bar{c}=0,1,2$. Here, $ H^{(3)\parallel} $ denotes longitudinal components of the three form flux and $ H^{(3)\perp} $ are the transverse world-volume components of the three form flux.

We denote the longitudinal three form on the world-volume as
\begin{align}
\label{2.22}
\mathcal{T}^{(3)}&=\mathcal{T}_{\bar{a} \bar{b}\bar{c}}(x) d \sigma^{\bar{a}} \wedge d \sigma^{\bar{b}} \wedge d \sigma^{\bar{c}} + \cdots \nonumber\\
&=\frac{1}{3!} \tau_{\bar{a}}~^{A_0}(x)\tau_{\bar{b}}~^{A_1}(x)\tau_{\bar{c}}~^{A_2}(x)\varepsilon_{A_0 A_1 A_2}d \sigma^{\bar{a}} \wedge d \sigma^{\bar{b}} \wedge d \sigma^{\bar{c}} + \cdots \nonumber\\
&=\mathcal{T}^{(3)}(x)+\mathcal{O}(c^{-2}),
\end{align}
where we introduce the pull-back of the longitudinal one forms as, $ \tau_{\bar{a}}~^{A}(x)=\tau_\mu~^A (x)\partial_{\bar{a}} x^\mu $.

Next, we calculate the following entity\footnote{The other component $H^{\ast \bar{a}\bar{b}\bar{c}}$ goes like $\sim c^{-3}$ which is therefore not important at LO.}
\begin{align}
\label{e28}
&H^{\ast ijk}=\frac{1}{3! \sqrt{- \det \hat{g}_{ab}}}\varepsilon^{ijk \bar{c}\bar{m}\bar{n}}H^{\parallel}_{\bar{c}\bar{m}\bar{n}}\nonumber\\
&=\frac{c^{3/2}}{3!}\frac{\mathcal{T}_{\bar{c}\bar{m}\bar{n}}(x)\varepsilon^{ijk \bar{c}\bar{m}\bar{n}}}{\sqrt{- \det \tau_{\bar{a}\bar{b}}^{(3)}(x)}\sqrt{\det h^{(3)}_{ij}(x)}}\Big(1+\frac{1}{2c^2}\frac{\Sigma(x,y)} {(-\det \tau_{\bar{a}\bar{b}}^{(3)}(x))}\Big)\nonumber\\
&+\frac{1}{\sqrt{c} 3!}\frac{y^\mu \partial_\mu \mathcal{T}_{\bar{c}\bar{m}\bar{n}}(x)\varepsilon^{ijk \bar{c}\bar{m}\bar{n}}}{\sqrt{- \det \tau_{\bar{a}\bar{b}}^{(3)}(x)}\sqrt{\det h^{(3)}_{ij}(x)}}+\cdots ,\nonumber\\
&\equiv c^{3/2}\Pi^{ijk}(x)+\frac{1}{\sqrt{c}}\tilde{\Pi}^{ijk} (x,y)+\cdots,
\end{align}
where we identify the function(s)
\begin{align}
\Pi^{ijk}(x)=\frac{1}{3!}\frac{\mathcal{T}_{\bar{c}\bar{m}\bar{n}}(x)\varepsilon^{ijk \bar{c}\bar{m}\bar{n}}}{\sqrt{- \det \tau_{\bar{a}\bar{b}}^{(3)}(x)}\sqrt{\det h^{(3)}_{ij}(x)}}.
\end{align}

Before we proceed further, it is worthwhile to identify inverse world-volume metric
\begin{align}
\label{e29}
&\hat{g}^{\bar{a}\bar{b}}=T^{\bar{a}\bar{b}}+c^{-2}\tau^{\bar{a}\bar{b}}~;~ \hat{g}^{ij}=c h^{ij}+\mathcal{O}(c^{-1})
\end{align}
such that the following orthonormality conditions hold true
\begin{align}
\tau_{\bar{a}\bar{b}}T^{\bar{b}\bar{c}}=0~;~\tau_{\bar{a}\bar{b}}\tau^{\bar{b}\bar{c}}+T_{\bar{a}\bar{b}}T^{\bar{b}\bar{c}}=\delta_{\bar{a}}~^{\bar{c}}~;~h_{ij}h^{jk}=\delta_{i}~^k.
\end{align}

Considering the expansions \eqref{e28} and \eqref{e29} together with the assumption that the auxiliary scalars ($ \phi = \phi (\sigma^i) $) are functions of transverse world-volume directions only, it is straightforward to compute
\begin{align}
\label{e32}
&-\text{tr}H^2 =H^{\ast}_{ ab}H^{\ast ab}=\frac{\hat{g}_{am}\hat{g}_{bn}}{(\partial \phi )^2 }H^{\ast mnc}H^{\ast abl}\partial_c \phi \partial_l \phi=\frac{h_{ij}h_{kl}}{(\hat{\partial \phi})^2}\Pi^{jlp}\Pi^{ikr}\partial_p \phi \partial_r \phi + \mathcal{O}(c^{-2}),
\end{align}
where we use the fact that under NR scaling
\begin{align}
&(\partial \phi )^2=c\partial_i \phi  h^{ij} \partial_j \phi = c (\hat{\partial \phi})^2.
\end{align}

On a similar note, one finds
\begin{align}
\label{ne34}
&(\text{tr}H^2)^2 = \frac{h_{i_1j_1}h_{k_1l_1}h_{i_2j_2}h_{k_2l_2}}{(\hat{\partial \phi})^4}\Pi^{j_1l_1p_1}\Pi^{i_1k_1r_1}\Pi^{j_2l_2p_2}\Pi^{i_2k_2r_2}\partial_{p_1} \phi \partial_{r_1} \phi \partial_{p_2} \phi \partial_{r_2} \phi + \mathcal{O}(c^{-2})\\
\label{ne35}
&\text{tr}H^4 =-\frac{g_{ap}g_{mq}g_{bs}g_{nl}}{(\partial \phi)^4}H^{\ast pqc}H^{\ast msd}H^{\ast ane}H^{\ast lbf}\partial_c \phi \partial_d \phi \partial_e \phi \partial_f \phi\nonumber\\
&= -\frac{h_{i_1 j_1}h_{i_2 j_2}h_{i_3 j_3}h_{i_4 j_4}}{(\hat{\partial \phi})^4}\Pi^{j_1j_2 r_1}\Pi^{i_2 j_3 r_2}\Pi^{i_1 i_4 r_3}\Pi^{j_4 i_3 r_4}\partial_{r_1}\phi \partial_{r_2}\phi \partial_{r_3}\phi \partial_{r_4}\phi +\mathcal{O}(c^{-2}).
\end{align}

Combining \eqref{3.12}, \eqref{e32}, \eqref{ne34} and \eqref{ne35},  one finds\footnote{Notice that, the $ c^{3/2} $ scaling is an artefact of the M5 brane embedding we choose to work with. Each longitudinal component of the metric contributes a factor of $ c^2 $ to the determinant while the transverse metric components ($ \hat{g}_{ij} $) come along with a factor of $ c^{-1} $. While taking the determinant, they produce an over all factor of $ c^3 $ under the square root.}
\begin{align}
\label{e34}
&\sqrt{-\det \hat{g}_{ab} }\left(1+\frac{1}{2}\text{tr}H^2 +\frac{1}{8} (\text{tr}H^2)^2 -\frac{1}{4}\text{tr}H^4\right)^{1/2}\nonumber\\
&=c^{3/2}\sqrt{-\det \tau_{\bar{a}\bar{b}}^{(3)}(x)}\sqrt{\det h^{(3)}_{ij}(x)}\Big(1-\frac{1}{2}\frac{h_{ij}h_{kl}}{(\hat{\partial \phi})^2}\Pi^{jlp}\Pi^{ikr}\partial_p \phi \partial_r \phi \nonumber\\
&+\frac{1}{8}\frac{h_{i_1j_1}h_{k_1l_1}h_{i_2j_2}h_{k_2l_2}}{(\hat{\partial \phi})^4}\Pi^{j_1l_1p_1}\Pi^{i_1k_1r_1}\Pi^{j_2l_2p_2}\Pi^{i_2k_2r_2}\partial_{p_1} \phi \partial_{r_1} \phi \partial_{p_2} \phi \partial_{r_2} \phi\nonumber\\
&+\frac{1}{4}\frac{h_{i_1 j_1}h_{i_2 j_2}h_{i_3 j_3}h_{i_4 j_4}}{(\hat{\partial \phi})^4}\Pi^{j_1j_2 r_1}\Pi^{i_2 j_3 r_2}\Pi^{i_1 i_4 r_3}\Pi^{j_4 i_3 r_4}\partial_{r_1}\phi \partial_{r_2}\phi \partial_{r_3}\phi \partial_{r_4}\phi\Big)^{1/2}+\mathcal{O}(1/c^{7/2}),
\end{align}
where $ \Pi^{ijk} $ are the transverse world-volume fluctuations of the M5 brane in the NR limit.

Next, we consider the NR expansion of $ H^{\ast }H (\partial \phi)^2 $ corrections pertaining to the kinetic energy action (\ref{2.1}). A straightforward computation yields
\begin{align}
\label{e35}
&\frac{\sqrt{-\det \hat{g}_{ab}}}{4 \partial_a \phi \partial^a \phi}\partial_l \phi H^{\ast lij}H_{ijp}\partial^p \phi \nonumber\\
&=\frac{c^{3/2}\sqrt{-\det \tau_{\bar{a}\bar{b}}^{(3)}(x)}\sqrt{\det h^{(3)}_{ij}(x)}}{4h^{ij}\partial_i \phi \partial_j \phi}h^{pr}\partial_l \phi \Pi^{lij}\pi_{ijp}\partial_r \phi + \cdots.
\end{align}

Combining \eqref{e34} and \eqref{e35}, we obtain the LO Lagrangian in the NR expansion
\begin{align}
\label{e36}
&L^{(kin)}_{LO}=\sqrt{-\det \tau_{\bar{a}\bar{b}}^{(3)}(x)}\sqrt{\det h^{(3)}_{ij}(x)}\Big[\Big(1-\frac{1}{2}\frac{h_{ij}h_{kl}}{(\hat{\partial \phi})^2}\Pi^{jlp}\Pi^{ikr}\partial_p \phi \partial_r \phi \nonumber\\
&+\frac{1}{8}\frac{h_{i_1j_1}h_{k_1l_1}h_{i_2j_2}h_{k_2l_2}}{(\hat{\partial \phi})^4}\Pi^{j_1l_1p_1}\Pi^{i_1k_1r_1}\Pi^{j_2l_2p_2}\Pi^{i_2k_2r_2}\partial_{p_1} \phi \partial_{r_1} \phi \partial_{p_2} \phi \partial_{r_2} \phi\nonumber\\
&+\frac{1}{4}\frac{h_{i_1 j_1}h_{i_2 j_2}h_{i_3 j_3}h_{i_4 j_4}}{(\hat{\partial \phi})^4}\Pi^{j_1j_2 r_1}\Pi^{i_2 j_3 r_2}\Pi^{i_1 i_4 r_3}\Pi^{j_4 i_3 r_4}\partial_{r_1}\phi \partial_{r_2}\phi \partial_{r_3}\phi \partial_{r_4}\phi\Big)^{1/2}\nonumber\\
&-\frac{1}{4(\hat{\partial \phi})^2}h^{pr}\partial_l \phi \Pi^{lij}\pi_{ijp}\partial_r \phi\Big],
\end{align}
which is subjected to the scaling, $ T_{NR}=c^{3/2}T_{M5} $ of the M5 brane tension in the NR limit. 

The LO theory \eqref{e36} is invariant under longitudinal and transverse diffeomorphisms. The theory also enjoys a global $SO(1,2)$ invariance (that rotates longitudinal $A$ and $B$ indices) along with a transverse $SO(8)$ invariance that rotates transverse indices, see \eqref{10}.
\subsection{Expansion of WZ term}
We now focus on the expansion of the WZ term \eqref{e2}. We propose the following expansions for the three form fluxes
\begin{align}
\label{e37}
&F_{ijk}=c^{-3/2}f_{ijk}(x)+\mathcal{O}(c^{-7/2})\\
&\hat{C}^{\parallel}_{\bar{a}\bar{b}\bar{c}}=c^3\mathcal{T}_{\bar{a}\bar{b}\bar{c}}(x)+\mathcal{O}(c)\\
&\hat{C}^{\perp}_{ijk}=c^{-3/2}\varpi_{ijk}(x)+\mathcal{O}(c^{-7/2}),
\label{e38}
\end{align}
where $ \hat{C}^{(3)\parallel} $ and $ \hat{C}^{(3)\perp} $ are respectively the parallel and transverse components of the world-volume pull back of the three form flux in the NR limit.

Combining \eqref{e37}-\eqref{e38}, one finds
\begin{align}
&F^{(3)}\wedge \hat{C}^{(3)}=\frac{c^{3/2}}{(3!)^2}f_{ijk}\mathcal{T}_{\bar{a}\bar{b}\bar{c}} d\sigma^i \wedge d\sigma^j \wedge d\sigma^k \wedge  d\sigma^{\bar{a}}\wedge  d\sigma^{\bar{b}}\wedge  d\sigma^{\bar{c}}+\cdots\nonumber\\
&=c^{3/2}f^{(3)}(x)\wedge \mathcal{T}^{(3)}(x)+\cdots,
\end{align}
where $ f^{(3)}=db^{(2)} $ for some antisymmetric two form field in the NR limit of M5 brane.

Finally, the world-volume pull back of the six form flux could be expressed as
\begin{align}
\hat{C}^{(6)}=c^{3/2}\varpi^{(3)}(x)\wedge \mathcal{T}^{(3)}(x)+\cdots.
\end{align}

Following the rescaling of the M5 brane tension, the LO WZ contribution reads as
\begin{align}
L_{LO}^{(wz)}=\varpi^{(3)}(x)\wedge \mathcal{T}^{(3)}(x)+\frac{1}{2}f^{(3)}(x)\wedge \mathcal{T}^{(3)}(x).
\end{align}
\section{Summary and final remarks}
Let us summarise the main result of this paper. The purpose of the present paper was to construct a NR covariant action for M5 branes based on a codimension three foliation \cite{Blair:2021waq} of the $ D=11 $ supergravity background. We consider a particular embedding of the M5 brane, where three of its world-volume directions are stretched along the longitudinal axes of the non-Lorentzian target space. As a result, the M5 brane theory effectively boils down into an infinitely extended three dimensional membrane, such that any field excitations along transverse axes should be treated as fluctuations of this extended object. This is precisely what we notice in \eqref{e36} where $ i,j,k $ are the indices along the transverse axes of the brane. It would be nice to extend the above ideas in the presence of superspace formalism of \cite{Bandos:1997ui} in which the $ D=11 $ background is replaced with the target space superfields that comprise of both the bosonic ($ X^\mu $) and the Grassmann directions ($ \Theta^{\mu} $).
\paragraph{Acknowledgements.}
The author is indebted to the authorities of IIT Roorkee for their unconditional support towards researches in basic sciences. The author also acknowledges The Royal Society, UK for financial assistance. The author would also like to acknowledge the Mathematical Research Impact Centric Support (MATRICS) grant (MTR/2023/000005) received from SERB, India.

\end{document}